\documentclass[a4paper]{spie}
\usepackage{graphicx}
\usepackage{bm}
\title{Numerical simulations of a quantum algorithm for Hilbert's tenth problem}

\author{Tien D Kieu
\skiplinehalf
Centre for Atom Optics and Ultrafast Spectroscopy,\\
Swinburne University of Technology, Hawthorn 3122, Australia}

\authorinfo{kieu@swin.edu.au}
\begin{document}
\maketitle
\begin{abstract}
We employ quantum mechanical principles in the computability exploration of the 
class of classically noncomputable Hilbert's tenth problem which is equivalent 
to the Turing halting problem in Computer Science.  The Quantum Adiabatic Theorem 
enables us to establish a connection between the solution for this 
class of problems and the asymptotic behaviour of solutions of a particular type of 
time-dependent Schr\"odinger equations.  We then present some preliminary numerical 
simulation results for the quantum adiabatic processes corresponding to various 
Diophantine equations.
\end{abstract}

%\pacs{}
%\keywords{}


\section{Hilbert's tenth problem}
One of the well-known and fascinating mathematical problems is about
the existence of integer solutions of polynomial equations
of the type 
\begin{eqnarray}
(x+1)^3 + (y+1)^3 - (z+1)^3 &\stackrel{?}{=}& 0. \nonumber
\end{eqnarray}
According to Fermat's last theorem, which has only been proved
very recently, the above equation does not have any integer
solution for $x$, $y$, and $z$.  However, although it took hundreds of years 
for mathematicians to finally come up with such a proof, this proof may not be 
valid for similar equations such as
\begin{eqnarray}
(x+1)^3 + (y+1)^3 - (z+1)^3 + xyz &\stackrel{?}{=}& 0. \nonumber
\end{eqnarray}
Many important mathematical problems, such as Goldbach's conjecture or
the distribution of zeroes of Riemann Zeta function, can be cast into 
equivalent problems of whether related polynomial
equations with integer coefficients have integer solutions or not.
Because of this, David Hilbert in his
famous list of mathematical problems in 1900 included as the tenth problem~\cite{tenth}
the challenge:
\begin{quotation}
Given any polynomial equation with any number of unknowns and with integer
coefficients:  To devise a universal process according to which it can be
determined by a finite number of operations whether the equation has
integer solutions.
\end{quotation}
Such equations are known as Diophantine equations.  Never was it anticipated
that this tenth problem is ultimately equivalent (as shown by Davis, Putnam, Robinson
and Matiyasevich, see~\cite{tenth}) to the halting problem of Turing machines of 
more than 30 years later.  
On this equivalence basis it has been concluded that Hilbert's tenth is not computable: 
there is no 
{\em single} universal process to determine the existence of integer solution or 
lack of it for arbitrarily given Diophantine equations in as far as there is no 
{\em single} universal machine to determine the halting or not of arbitrarily given 
Turing machine (which starts with some arbitrarily given input).  

Thus, we would have to consider 
anew each different case of Diophantine equation.

In spite of this widely accepted result, we have proposed a
{\em quantum algorithm}~\cite{kieu1} for solving Hilbert's tenth problem.  In the next
section we briefly summarise the algorithm.  We then next present some preliminary
numerical results from the simulations of quantum processes for some very simple
Diophantine equations as a concrete ``proof of concept".

\section{A quantum algorithm}
\subsection{General oracle for Hilbert's tenth problem}
It suffices to consider only nonnegative solutions of a Diophantine equation.
Let us consider the example
\begin{eqnarray}
(x+1)^3 + (y+1)^3 - (z+1)^3 + cxyz = 0, && c\in Z,
\nonumber
\end{eqnarray}
with unknowns $x$, $y$, and $z$.  Starting from the observation that if
we can construct the hamiltonian
\begin{eqnarray}
H_P &=& \left((a^\dagger_x a_x+1)^3 + (a^\dagger_y a_y+1)^3 - (a^\dagger_z
a_z+1)^3 + c(a^\dagger_x a_x)(a^\dagger_y a_y)(a^\dagger_z a_z) \right)^2,
\nonumber
\end{eqnarray}
which has a spectrum bounded from below in fact, and if we can
obtain the corresponding ground state (of least energy) then we can solve the tenth
problem!

The ground state $|g\rangle$ 
of the hamiltonian so constructed has the properties,
for some $(n_x,n_y,n_z)$,
\begin{eqnarray}
N_j|g\rangle &=& n_j|g\rangle, \nonumber\\
H_P|g\rangle &=& \left((n_x+1)^3 + (n_y+1)^3 - (n_z+1)^3 + cn_xn_yn_z
\right)^2|g\rangle \equiv E_g |g\rangle.
\nonumber
\label{eigenvalues}
\end{eqnarray}
Thus a projective measurement of the energy $E_g$  of the ground state 
$|g\rangle$ will yield 
the answer for the decision problem:
{\it\bf\it The corresponding Diophantine equation has at least one
integer solution if and only if $E_g = 0$, and has not otherwise.}
(If $c=0$ in our example, we know that $E_g > 0$ from the Fermat's last theorem.)

%\noindent
%\framebox[9.5 in]{\parbox{9in}{
From the above observation, our
ground-state oracle is thus clear:
\begin{enumerate}
\item  Given a Diophantine equation with $K$ unknowns $x$'s
\begin{eqnarray}
D(x_1,\cdots,x_K) &=& 0,
\end{eqnarray}
we need to simulate on some appropriate Fock space
the quantum hamiltonian 
\begin{eqnarray}
H_P &=& \left(D(a^\dagger_1 a_1,\cdots, a^\dagger_K a_K) \right)^2.
\end{eqnarray}
\item  Measurement results of appropriate observables in the ground state 
will provide the answer for our decision problem.
\end{enumerate}

One way, which is by no mean the only way, to obtain the ground state is
guaranteed by the quantum adiabatic theorem~\cite{messiah} which we will 
exploit in the next section.

\subsection{Quantum Adiabatic Algorithm}
In the adiabatic approach~\cite{mit}, one
starts with a hamiltonian $H_I$ whose ground state $|g_I\rangle$ is 
readily achievable.  Then one forms a ``slowly" varying hamiltonian
which interpolates between $H_I$ and $H_P$ in the time interval $t\in[0,T]$
\begin{eqnarray}
{\cal H}(t/T) &=& \left(1-t/T\right)H_I + 
(t/T) H_P,\nonumber\\
&\equiv& H_I + (t/T)W.
\end{eqnarray}
We will adopt this approach with the proposed (universal) initial Hamiltonian
\begin{eqnarray}
H_I &=& \sum_{i=1}^K (a^\dagger_i - \alpha^*_i)(a_i-\alpha_i)
\end{eqnarray}
which admits as the ground state the coherence state 
\begin{eqnarray}
|g_I\rangle = |\alpha_1\rangle \otimes \ldots |\alpha_K\rangle,
\label{cohstate}
\end{eqnarray}
where $|\alpha_i\rangle = e^{-\frac{1}{2}|\alpha_i|^2}
\sum_{n=0}^\infty \frac{\alpha_i^n}{\sqrt{n!}}|n\rangle.$

Provided the conditions of the adiabatic theorem~\cite{messiah} are observed,
the initial ground state will evolve into our
desirable ground state $|g\rangle$ up to a phase:
\begin{eqnarray}
\lim_{T\to\infty}{\cal T}\exp\left\{ -iT\int_0^1 {\cal H}(\tau) 
d\tau\right\}|g_I\rangle &=& {\rm e}^{i\phi}|g\rangle,
\label{7}
\end{eqnarray}
where $\cal T$ is the time-ordering operator.

%\section{The Implementation}
So our problem now is to solve the time-dependent Schr\"odinger equation
for $t\in [0,T]$
\begin{eqnarray}
i\partial_t \psi(t) &=& {\cal H}(t/T) \psi(t),\; t\in[0,T],
\end{eqnarray}
with the initial state being the coherence state~(\ref{cohstate}).

We can analytically show in general the two crucial results below~\cite{kieu_prep}:
\begin{itemize}
\item The ground state of ${\cal H}(t/T)$ is non-degenerate for $t\in(0,T)$.
As the minimum energy gap between the ground state and the first excited state
is non-zero, it takes only a
finite time $T$ for the adiabatic process, as asserted by the quantum adiabatic theorem,
to generate a state which has a high probability  of being the ground state of $H_P$.

\item The probability of the state at time $T$ in some number state, $|\langle
\psi(T)|n_0\rangle|^2$, is greater than 1/2 {\em iff} $|n_0\rangle$ is the ground state 
of $H_P$.  That is,
we only need to solve the equation for increasing $T$ until this majority condition is 
satisfied in order to {\em identify} the ground state.
\end{itemize}
The proofs of these important results will be available elsewhere.

\section{Simulation Technicalities}
\begin{itemize}
\item We solve the Schr\"odinger equation numerically in some finitely truncated 
Fock space large
enough to approximate the initial coherence state to an arbitrarily given accuracy.
That is, a truncation $m$ is chosen such that
\begin{eqnarray}
|\alpha_i;m\rangle = e^{-\frac{1}{2}|\alpha_i|^2}
\sum_{n=0}^m \frac{\alpha_i^n}{\sqrt{n!}}|n\rangle
\end{eqnarray}
has a norm less than one by some chosen $\epsilon$,
$|1- \sqrt{\langle \alpha_i;m|\alpha_{i};m\rangle} | \le\epsilon$. 

\item At each time step $\delta t$, and up to $O(\delta t^2)$,
\begin{eqnarray}
\psi(t+\delta t) &=& exp\{-i{\cal H}(t/T)\delta t\} \psi(t) = 
(1 -i{\cal H}(t/T)\delta t -1/2 {\cal H}^2(t/T)\delta t^2)\psi(t),
\end{eqnarray}
thus there are maximally only two creation operators, $a^\dagger_i a^\dagger_j$, and
we can explore this fact to {\it
explore the infinite Fock space} by increasing the size of the truncated Fock space by two
at every time step.

\item We employ the unitary solver, with conjugate gradient method,
\begin{eqnarray}
\psi(t+\delta t) &=& \frac{1-\frac{i}{2}{\cal H}(t/T)\delta t}
{1 + \frac{i}{2}{\cal H}(t/T)\delta t} \psi(t)
\end{eqnarray}
which approximates $exp\{-i{\cal H}(t/T)\delta t\}$ to second order in $\delta t$.
This solver, being unitary, preserves the norm of the state vector in the evolution
of time even though the size of the underlying truncated Fock space is allowed to
increase with time if necessary.

\item The time step $\delta t$ at time $t$ is a function of time and 
is chosen in such a way that halving this $\delta t(t)$
only results in $O(\delta t^2)$ correction.
\end{itemize}

\section{Simulation parameters and results}
\subsection{Equation $xy+x+4y-11=0$}
This is an equation resulted from the factoring of the number 15 into two
prime factors.  Because of the nature of the problem we can fix the truncated
size of our Fock space to be $(m_x,m_y)=(9,9)$ in order to have the norm of the 
state vector less than unity by an amount $\epsilon = 10^{-2}$ at all times.  
In this and in all other simulations
below we choose $\alpha_x = \alpha_y = (2.0,0.0)$.

\begin{figure}
\begin{center}
\includegraphics{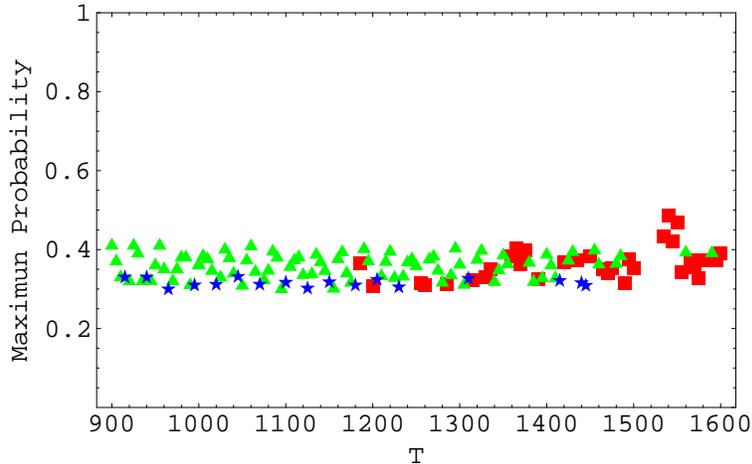}
\caption{\label{Fig1}Corresponding equation $xy+x+4y-11=0$; with intermediate
evolution time $T$.}
\end{center}
\end{figure}

In Fig.~\ref{Fig1} we plot the magnitude square of maximum component (in terms
of the Fock states) of the state 
vector as a function of the total evolution time $T$ in some arbitrary unit.
Closed to below $T\sim 1500$, the maximum probability components of $|\psi(T)\rangle$ 
are dominant by some states, denoted by (blue) star and (green) triangle symbols, 
none of which are the ground state.  In fact, they are, respectively,
the {\em first} degenerate excited states $|n_x, n_y\rangle = |4,1\rangle$ and $|3,1\rangle$ 
of $H_P$.

\begin{figure}
\begin{center}
\includegraphics{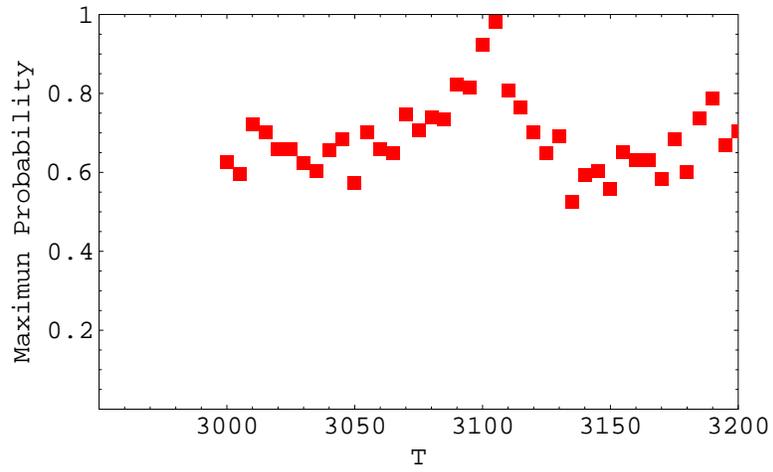}
\caption{\label{Fig2}Corresponding equation $xy+x+4y-11=0$; with $T$ entering the
quantum adiabatic regime.}
\end{center}
\end{figure}

With the increase in $T$ in Fig.~\ref{Fig2}, one of the Fock state, $|1,2\rangle$ 
denoted by (red) box symbol, has probability greater than $\frac{1}{2}$
which is our criterion for being identified as the ground state.  We mark this regime as
the quantum adiabatic regime when the ground state wins the battle for dominance.
From the ground state so identified we can infer that our Diophantine equation has
one solution in this domain.

\subsection{Equation $x+20 = 0$}
We consider this extremely simple equation as an example which has no solution 
in the positive integers.  The simulation parameters are as in the previous, except
that the truncated size of our Fock space, starting with size 8, is now allowed to vary
with time in order to simulate the exploration of the whole infinite space.

\begin{figure}
\begin{center}
\includegraphics{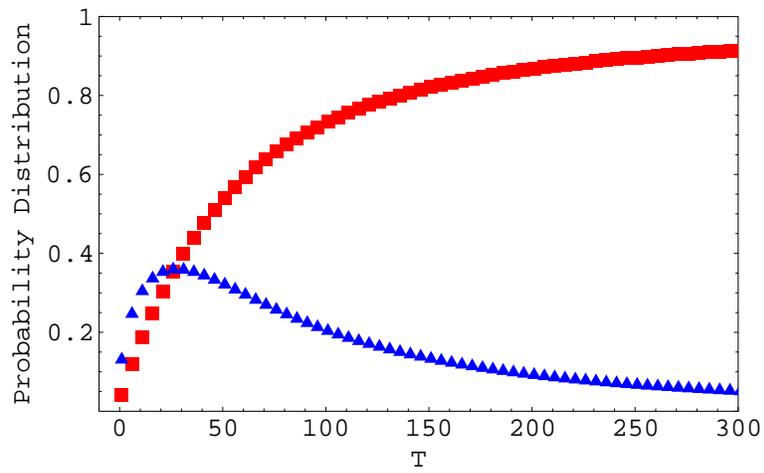}
\caption{\label{Fig3}Corresponding equation $x+20=0$.}
\end{center}
\end{figure}

In Fig.~\ref{Fig3} we plot the probabilities of the dominant components as a function
of $T$.  Below $T\sim 50$ none of the two components is greater than one-half, and in fact
the first excited state, $|1\rangle$ denoted by (blue) triangle symbol, clearly dominates 
in this regime.
Eventually we enter the quantum adiabatic regime upon when the (red) box symbol rises
over the one-half mark; indeed it corresponds to the Fock state $|0\rangle$ which
is the true ground state and which implies that our original Diophantine equation has no
integer solution at all.

\subsection{Equation $x-20 = 0$}
We now consider a simple example which nevertheless has all the interesting ingredients
typical for a general simulation of our quantum algorithm.  Here we choose
$\epsilon = 10^{-3}$ and the initial Fock space has only up 
to $|n\rangle = |14\rangle$, which does not include the true ground state of $H_P$.  
This is typical in our simulations since we in general would not be able to tell
in advance
whether our initial Fock spaces do contain the true ground states or not.  Generally,
they do not.  Even so, our strategy of allowing the expansion in the size of truncated
Fock space in time has enabled the true ground state to be found and identified.

\begin{figure}
\begin{center}
\includegraphics{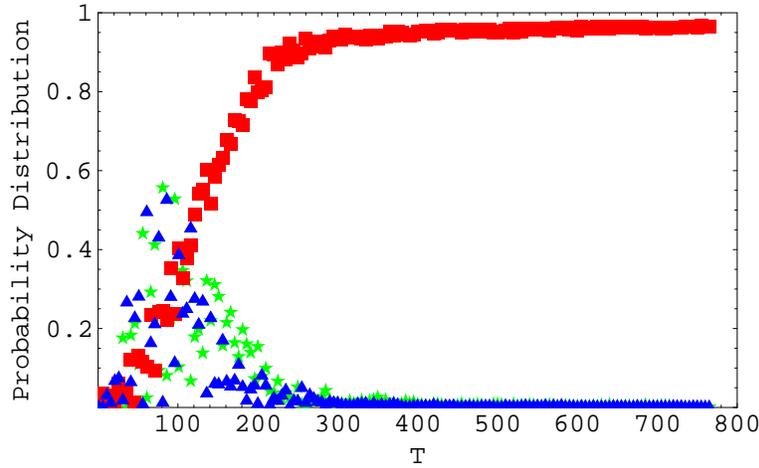}%
\caption{\label{Fig4}Corresponding equation $x-20=0$.}
\end{center}
\end{figure}

In this example, the state $|20\rangle$, {\em which is not included in the initial truncated
Fock space}, is eventually reached and identified as the ground state 
as shown as red boxes in Fig.~\ref{Fig4}.  Blue triangles  and green stars are corresponding 
to the first two excited states $|19\rangle$ and $|21\rangle$, which are degenerate
eigenstates of $H_P$.

Note that these competing pretenders somehow have unexpectedly 
probabilities greater than one-half (around $T\sim 90$), contrary to our analytical
result that only the ground state can have probability rising above one-half!  We think that
this is only some artefact of finite-size time steps $\delta t$, and expect that
it would go away once we employ a more sophisticated method for solving the 
Schr\"odinger equation. 
Work is in progress to systematically extrapolate to zero-size time steps to confirm the
removal of this type of finite-size effects.

%\subsubsection{}

% If in two-column mode, this environment will change to single-column
% format so that long equations can be displayed. Use
%\begin{widetext}
%\end{widetext}

% figures should be put into the text as floats.
% Use the graphics or graphicx packages (distributed with LaTeX2e)
% and the \includegraphics macro defined in those packages.
% See the LaTeX Graphics Companion by Michel Goosens, Sebastian Rahtz,
% entire page. There is no need to do explicit centering.

\begin{figure}
\begin{center}
\includegraphics{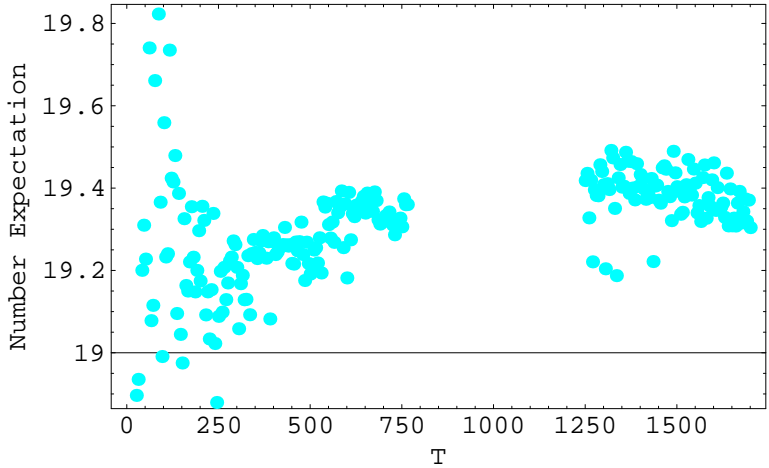}
\caption{\label{Fig5}The expectation value of occupation number, $\langle N\rangle$, 
as a function of $T$.}
\end{center}
\end{figure}

\begin{figure}
\begin{center}
\includegraphics{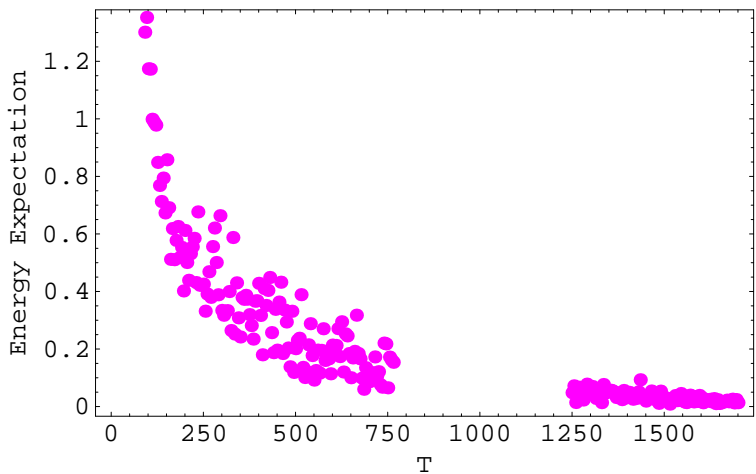}
\caption{\label{Fig6}The expectation value of energy, $\langle H_P\rangle$, 
as a function of $T$.}
\end{center}
\end{figure}

Figs.~\ref{Fig5} and~\ref{Fig6} depict the expectation values of occupation number
and of energy as functions of $T$.  They are, respectively, approaching 20 and zero
(which signify the fact that our equation has an integer solution, namely 20).

\section{Concluding Remarks}
We here consider the 
issue of computability in principle, not that of computational complexity.  
The simulation results support the idea of implementation and
execution of our quantum algorithm with a {\em physical} process~\cite{kieu1}:
\begin{itemize}
\item Run the physical process (corresponding to the
Diophantine equation in consideration) for some time $T$. 
% which might be estimated by the Bogoliubov transformations.
\item Repeat the process at this $T$ to obtain the statistics through measurements.
\item If none of the measurement outcomes exhibits probability of more than 1/2, 
ram up the $T$ and go back to the step above. 
\item Eventually, at some sufficiently large $T$, the measurement state with 
more-than-even probability can thus be identified as the ground state, terminating
our physical implementation and execution.
\item Substituting the quantum numbers for the now identified ground state enables
us to see if $E_g=0$ and thus whether the Diophantine equation has a solution.
\end{itemize}

% From the simulations herein arise the questions:  Is it possible to simulate
% the algorithm on classical computers for arbitrarily given Diophantine equation?
% How does this possibility relate to the well-known result of non-decidability 
% on classical computers of Hilbert's tenth problem?
% We are investigating these questions to identify which of the below:
% 
% Here, our non-computability components are:
% \begin{itemize}
% \item The number states $|n\rangle$, which are the eigenstates of $H_P$, are known {\em 
% a priori} and thus need not be computable in the classical sense.  
% (Pour-El and Richards~\cite{pour-el} have shown that for linear 
% operators on Banach space in general, or on Hilbert space in particular, their 
% eigenvalues are computable but their eigenvectors are {\bf not}.  Here, the
% eigenvectors $|n\rangle$, in contrast, are already given at the outset.)
% \item Our proof of the criterion for ground state identification is 
% mathematically {\it non-constructive}
% and thus need not be computable in the classical sense.
% \item The exponential behaviour onset when used as criterion is not computable in general.
% \end{itemize}
% or some other element of our quantum algorithm constitutes the crucial non-computability 
% element that allows its simulation on classical computers.
% \section*{Acknowledgments}
\begin{acknowledgments}
I would like to thank Cristian Calude, Bryan Dalton, Peter Hannaford, Alan Head, 
Toby Ord and Andrew Rawlinson 
for discussions and support.  I am also grateful to Mathew Bailes for the extensive 
use of Swinburne Supercluster facility to produce the numerical results reported herein, 
and to Barbara McKinnon for discussions on the finer points of FORTRAN90.
\end{acknowledgments}

% \bibliography{quadiabatic.bib}

\end{document}